# Symmetry breakdown in franckeite: spontaneous strain, rippling and interlayer moiré

*Riccardo Frisenda[1, ‡], Gabriel Sanchez-Santolino[1,*] †,‡, Nikos Papadopoulos[2], Joanna Urban[3], Michal Baranowski[3,4], Alessandro Surrente[3], Duncan K. Maude[3], Mar Garcia-Hernandez[1], Herre S. J. van der Zant[2], Paulina Plochocka[3,*], Pablo San-Jose[1,*], Andres Castellanos-Gomez[1,*].*

[1] Materials Science Factory, Instituto de Ciencia de Materiales de Madrid (ICMM-CSIC), Campus de Cantoblanco, E28049 Madrid, Spain.
[2] Kavli Institute of Nanoscience, Delft University of Technology, Lorentzweg 1, Delft 2628 CJ, The Netherlands
[3] Laboratoire National des Champs Magnétiques Intenses, UPR 3228, CNRS-UGA-UPS-INSA, Grenoble and Toulouse, France.
[4] Department of Experimental Physics, Faculty of Fundamental Problems of Technology, Wroclaw University of Science and Technology, Wroclaw, Poland
† Present address at Department of Physics, Chalmers University of Technology, 412 96, Gothenburg, Sweden.
‡ These authors contributed equally to this work

* Riccardo.frisenda@csic.es , gabriel.sanchezsantolino@chalmers.se, paulina.plochocka@lncmi.cnrs.fr, pablo.sanjose@csic.es, Andres.castellanos@csic.es

**Abstract:** Franckeite is a naturally occurring layered mineral with a structure composed of alternating stacks of $SnS_2$-like and PbS-like layers. Although this superlattice is composed of a sequence of isotropic two-dimensional layers, it exhibits a spontaneous rippling that makes the material structurally anisotropic. We demonstrate that this rippling comes hand in hand with an inhomogeneous in-plane strain profile and anisotropic electrical, vibrational and optical properties. We argue that this symmetry breakdown results from a spatial modulation of the van der Waals interaction between layers due to the $SnS_2$-like and PbS-like lattices incommensurability.

## Introduction

Since the isolation of graphene and other 2D materials by mechanical exfoliation [1, 2] a large fraction of the community working on this family of materials turned their attention towards the possibility of fabricating artificial heterostructures and superlattices by stacking dissimilar 2D materials on top of each other.[3-8]

These stacks can show electronic and optical properties that strongly differ from those of the constituent 2D materials, thus opening the door to the fabrication of materials with user-designed properties. The most extended method to fabricate 2D heterostructures and superlattices consists on the manual or robotic assembly of the stacks using deterministic placement methods.[9-12] Alternatively to those top down approaches, it has been recently shown that van der Waals superlattices can be naturally found in certain layered minerals, like the sulfosalts franckeite and cylindrite, that overcome a thermodynamic phase separation during their growth.[13-16] In fact, franckeite and cylindrite are layered minerals with a structure composed of alternating stacks of $SnS_2$-like pseudo-hexagonal (H) and PbS-like pseudo-tetragonal (Q) layers.[17] Franckeite has been exfoliated (mechanically and by liquid phase exfoliation) down to the single unit cell and the exfoliated flakes have been assembled into electronic devices and photodetectors operating in the near-infrared range.[13, 14, 18-21]

Here we present a study of an intriguing feature of franckeite: although the crystal is composed of isotropic 2D layers, the crystal exhibits a spontaneous rippling that makes the material structurally anisotropic. We show that rippling comes together with an inhomogeneous in-plane strain profile and anisotropic electrical and optical properties. Using a simple theoretical model, we show that this symmetry breakdown results from a spatial modulation of the van der Waals interaction between layers due to H-Q lattice incommensurability. Franckeite is thus a clear example of a natural superlattice system where the interaction between the constituent stacked layers give rise to new properties (anisotropy) not present in the individual 2D constituents (isotropic).

## Results and discussions

Franckeite flakes have been isolated by mechanical exfoliation of bulk natural franckeite mineral (San José mine, Oruro, Bolivia) with Nitto tape and transferred to other substrates with Gel-film® (by Gel-pak) using an all-dry transfer technique.[22] We point the reader to Ref. [13] for a detailed characterization of the franckeite bulk crystal.

Figure 1a shows an optical microscopy image of a franckeite flake, whose crystal structure is represented in the inset of Fig. 1a, which has been transferred onto a holey $Si_3N_4$ transmission electron microscopy (TEM) grid. Figure 1b is a low magnification TEM image acquired on a freely-suspended part of the flake shown in Figure 1a, and 1c is the corresponding high-resolution image. Defocus was optimized to increase the contrast from a striped pattern which is superimposed on the atomic lattice and it can also be seen in the selected area diffraction pattern (SAED) in Figure 1d. The SAED exhibits the reflections of the H and Q layers along with rows of weak superlattice spots along the *c* direction due to the striped pattern. A direct comparison between panels (a) to (d) allows us to determine that the stripes in the franckeite flake are parallel to the flat edge shown in Figures 1a and 1b.

The striped pattern has been previously observed by Williams and Hyde and it has been found to arise from an out-of-plane deformation of the lattice, which develops periodic ripples with a period of ∼4.7 nm and a peak-to-peak amplitude of ∼0.08-0.14 nm out of plane along the franckeite c axis.[17, 23-25] The origin of this periodic rippling is still a subject of debate.

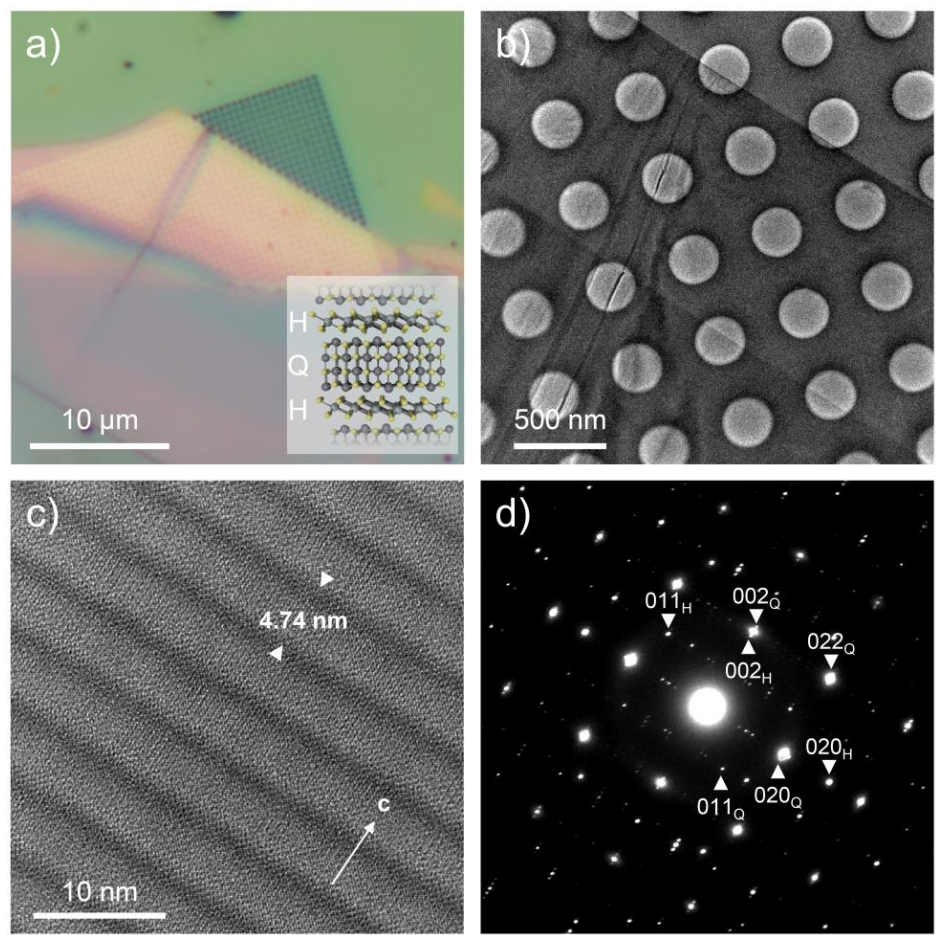

**Fig. 1. TEM characterization of the structure of franckeite superlattice.** (a) Optical image of a franckeite flake over a holey $Si_3N_4$ TEM support. Inset: crystal structure of franckeite. (b) Low magnification HRTEM image of the same franckeite flake. (c) Atomic resolution HRTEM image of the same flake down the (100) direction. (d) Selected area diffraction pattern (SAED) depicting the reflections of both alternating $SnS_2$-like pseudo-hexagonal (H) and PbS-like pseudo-tetragonal (Q) lattices.

Careful inspection of the structural deformations giving rise to the periodic striped pattern was carried out by strain analysis. Figure 2a shows a high magnification HRTEM image of a thin region of the franckeite flake in which we performed the analysis (see methods). We use the whole field of view within the analyzed image as reference for the strain analysis so the values obtained are expressed with respect to the mean lattice spacing of the analyzed region. The strain tensor $\varepsilon$ component along the c direction, shown in Fig. 2b, depicts strong alternating in-plane expansive and compressive strained regions, which are evidenced in the average profile in Fig. 2c. The periodicity of the spatially modulated in plane strain is 4.77 nm which is in good agreement with the period of the ripple pattern found by direct inspection on the HRTEM image in Fig. 1c. This finding suggests that the periodic rippling not only originates from an out of plane deformation of the lattice, as previously believed, but also from a strong in-plane strain modulation.

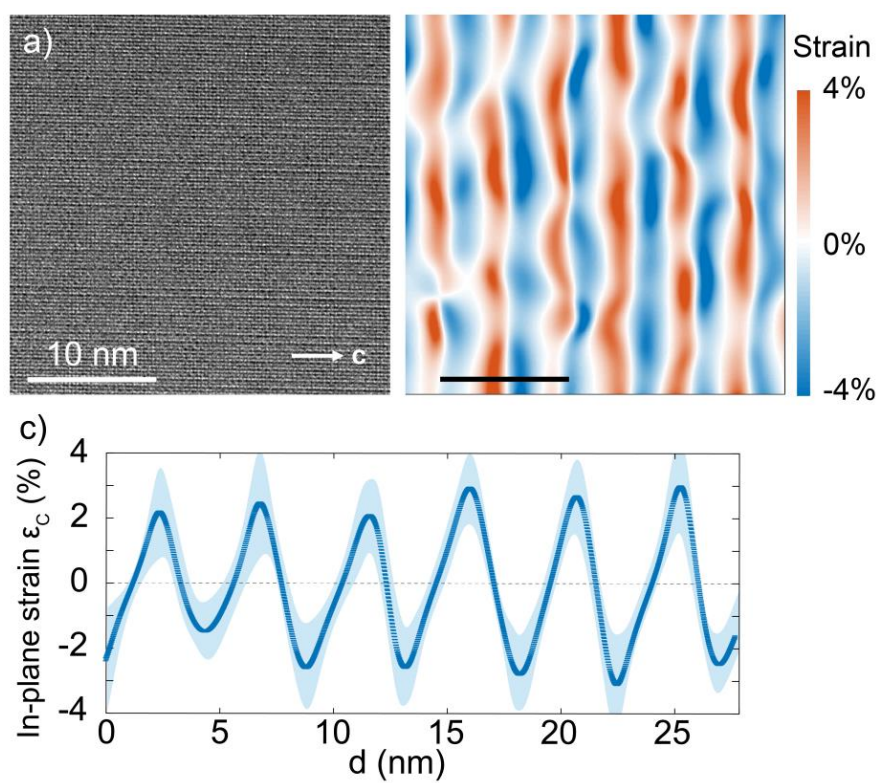

**Figure 2. Strain profile analysis of franckeite superlattice.** (A) HRTEM image of a franckeite flake down the (100) direction. (B) In-plane strain map along the rippling direction ($\varepsilon_c$ component of the strain tensor) obtained from the HRTEM image in (A) by geometric phase analysis (GPA). (C) Averaged in-plane strain $\varepsilon_c$ profile, calculated by integrating the strain map of panel (B) along the vertical direction, depicting the periodic expansive and compressive strain modulations along the rippling direction. The shaded area represents the scattering from the mean strain value given by the standard deviation.

In order to get a further insight into the origin of this striped pattern and the strong spatially modulated strain we have employed a theoretical model, developed in detail in the Supplementary Material, that de-scribes, within a continuum medium approximation, how rippling arises from the competition between in-homogeneous van der Waals interlayer interactions and inhomogeneous elastic deformations. The van der Waals adhesion energy between the H and Q layers is position dependent, following the interlayer moiré pattern, as it is controlled by the local atomic alignment between layers, which in turn depends on any inhomogeneous elastic deformations in the material. As the crystal relaxes mechanically to minimise the total energy (adhesion plus elastic) it develops both strong in-plane strains and a small out-of-plane rippling, modulated along the armchair direction of the H layer. The theoretical results are in quantitative agreement with the experimental observations. Figure 3a shows a typical lattice structure after relaxation. Figure 3b shows the adhesion potential between the hexagonal layers and the bottom/top tetragonal layers. Figures 3c and 3d show the profiles of ripples $h(y)$, in-plane deformations $u_y(y)$ and in-plane strain $\varepsilon_c(y)=du_y/dy$. The latter exhibits a period that is halved respect to the ripple profile, in agreement with our observations. These results clearly point to adhesion moiré patterns as the most likely origin for franckeite rippling and the observed strong in-plane strains. The relaxation of strain that induces a spatial modulation in the franckeite natural heterostructure belongs to a family of similar phenomena in van der Waals incommensurate lattices.[26-28] However, franckeite constitutes a unique case since the buckling instability is rather different from the corrugation effects in graphene/hBN and similar systems.

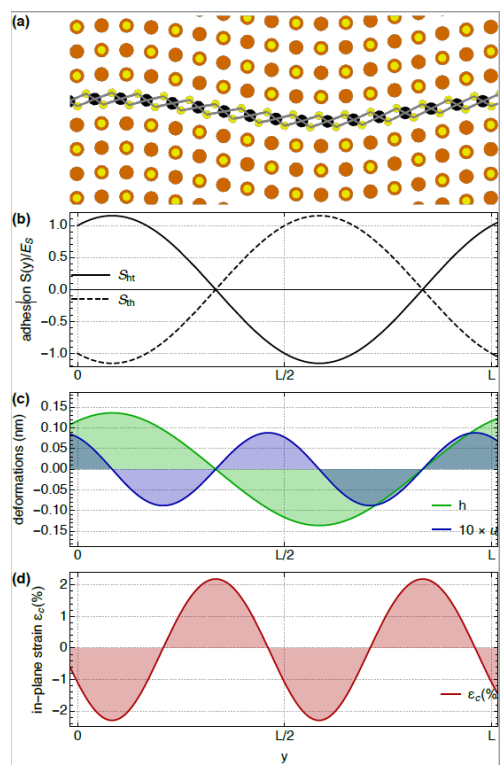

**Figure 3. Theoretical model to explain the symmetry breakdown in frankeite.** (A) Side view of the computed equilibrium structure of Franckeite across a ripple. with Pb in orange, S in yellow and Sn in black. The shear modulus G and the 2D Young modulus E are chosen so that $GV/LE_S=128$ and $EV/LE_S=45$, where $E_S = -E_{S'}$ is the maximal adhesion, V is the sample volume and L is the ripple period, See Supplementary Information for details on the theory. (B) Adhesion energy density between hexagonal layer and the tetragonal layers immediately below ($S_{ht}$) and above ($S_{th}$), normalized to the maximum adhesion $E_S$. (C) Equilibrium deformations out of plane (h(y), green) and in-plane ($u_y(y)$, blue). (D) Equilibrium in-plane strains $\varepsilon_c = du_y/dy$, which show variations around 4% peak-to-peak.

This periodic pattern of lattice strain can be the seed for further anisotropic optical and electrical properties. We first studied the anisotropy in the optical properties by measuring differential reflectance (a magnitude that is proportional to the absorption, see the Supporting Information) upon incidence with linearly polarized light.[29] Figure 4a shows some differential reflectance spectra, acquired for the same flake studied by TEM in Figure 1, as a function of the rotation angle of the linear polarizer. The spectra have been labelled with the angle formed between the striped pattern (determined from the TEM images) and the linearly polarized light (with 0º corresponding to linearly polarized light along the striped pattern and 90º linearly polarized light perpendicular to the stripes). Figure 4b shows the differential reflectance magnitude extracted at a fixed illumination wavelength (496 nm, 2.5 eV) as a function of the angle between the stripes and the linear polarization of the incident light. The data shows a clear angular dependence which is also evident from the polar plot in Figure 4c, indicating that franckeite shows linear dichroism (*i.e.* absorption dependent on the orientation of incident linearly polarized light) even though the individual H and Q layers are not expected to show this in-plane anisotropy. Apart from the periodical in-plane strain, which is known to strongly modulate the band-gap in 2D materials,[30-32] also the moiré pattern could induce some anisotropy through its effect on the electronic band structure. Nevertheless we do not expect an important contribution from the moiré modulation, since it modulates the electronic structure along both principal axes in the plane and thus it is not strongly anisotropic.

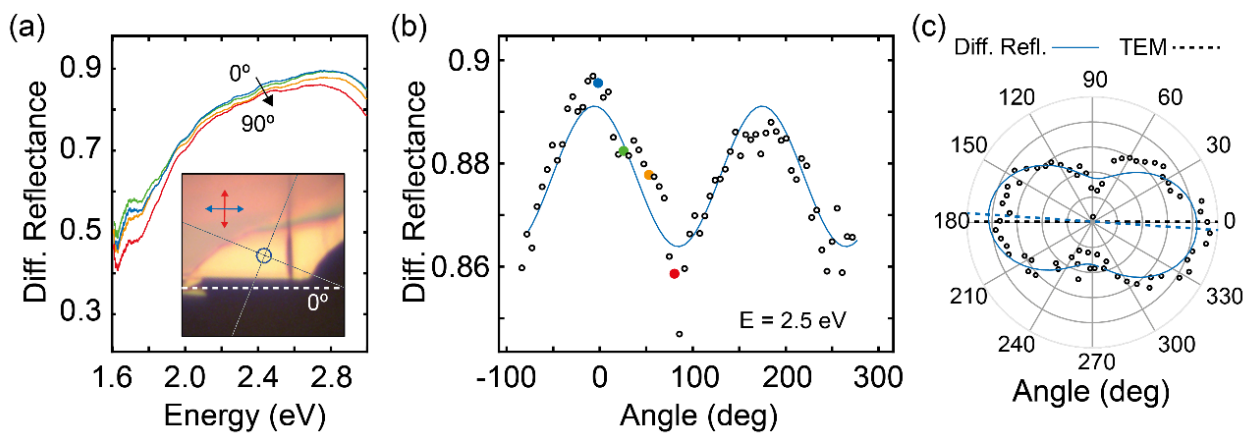

**Figure 4. Linear dichroism of frankeite.** (A) Differential reflectance spectrum recorded as a function of the rotation angle of the linear polarizer. Inset: optical image of the frankeite flake over a Gel-film substrate prior to its transfer to the TEM grid. (B) Differential reflectance at an energy of 2.5 eV (wavelength 496 nm) recorded as a function of the incident light linear polarization angle. (C) Same as (B) shown in polar coordinates. The dashed lines indicate the direction of the maximum of reflectance/absorbance (blue) and the direction of the ripples extracted from the HRTEM image of Figure 1 (black).

We further study the in-plane symmetry of the electronic properties by fabricating a nano-device with electrodes designed to address the electrical transport along different crystal orientations (see the optical microscopy image in Figure 5a and the topography of the device in Figure 5b). Figure 5c shows the conductance at different crystal orientations (note that the transport parallel to the long edge of the crystal has been labelled as 0º). The electrical conductivity varies by almost a factor of 2 from the transport at 0º (parallel to the long edge of the flake) to the transport at 90º (perpendicular to the long edge of the flake) and it follows a sinusoidal relationship that becomes even more evident in the elliptical shape of the data in the polar plot. In order to correlate the electrical measurements with the crystal structure we performed micro-reflectance measurements on the same frankeite flake before transferring it to the SiO$_2$/Si substrate in a similar way as discussed for Figure 4. Interestingly, from the micro-reflectance measurements we infer that the long edge of the flake shown in Figure 5a is parallel to the striped pattern of frankeite. Therefore one can conclude that the conductivity of the device is maximum along the crystal direction parallel to the stripes. Such an increase in conductance could be explained by an increase in the scattering rate experienced by the charge carriers travelling in the direction perpendicular to the ripples, which could results in a decrease of the mobility along that direction. The anisotropic electrical conductivity would also explain the observed linear dichroism: the flakes react to linearly polarized incident light similarly to a polaroid film or a wire-grid polarizer (incident light polarized parallel to the highly conductive crystal axis leads to a higher excitation of charge carriers and thus stronger dissipation of *E*-field).

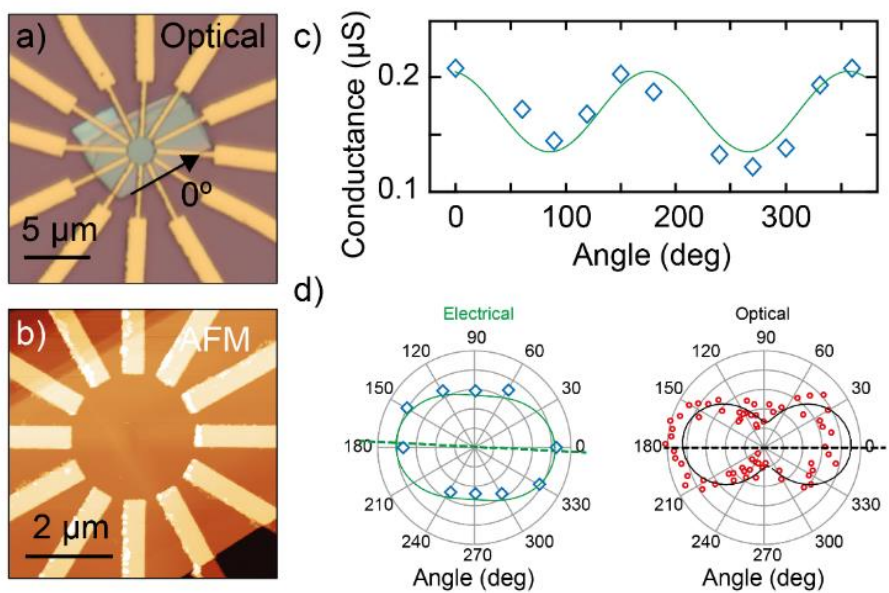

**Fig. 5. In-plane anisotropy of electrical properties of franckeite.** (A)-(B) Optical image (A) and AFM topography (B) of a franckeite flake over a SiO$_2$/Si substrate contacted with Au electrodes. (C) Conductance recorded between different pairs of electrodes. (D) Conductance (left) and differential reflectance recorded on a Gel-Film substrate prior its transfer to SiO$_2$/Si (right) of the franckeite flake shown in polar coordinates.

The in-plane anisotropy of the crystal is also naturally revealed by Raman spectroscopy performed using linearly polarized light. In this case, some of the vibrational modes exhibit strong increase of the intensity for the light polarized along one of the anisotropy axes. We have performed polarization resolved Raman measurements in the co-polarized configuration, namely for parallel polarization of the excitation and scattered light, and as a function of the angle between polarization axis and the orientation of the crystal. A typical set of polarized Raman spectra obtained at two different angles is shown in Figure 6a. Several peaks observed in our spectra are in agreement with previously reported data.[13, 14, 18] Peaks at 189 and 322 cm$^{-1}$ correspond to vibrational modes of the hexagonal layer of SnS$_2$.[33, 34] The peak at 263 cm$^{-1}$ (at polarization 0º) and the shoulder at 282 cm$^{-1}$ can be assigned to the vibrations of Sb$_2$S$_3$, however the energies of these modes are lower than in isolated stibnite.[35, 36] This difference is attributed to the complex nature of the tetragonal layer which contains also Pb and Sn atoms. Finally the low energy peak at 150 cm$^{-1}$ can be related to the vibrations of the PbS or SnS$_2$ lattices.[18] The angle dependence of the intensity of the Raman peaks in copolarized configuration (extracted from the fit of the Lorentzian function) is presented in Figure 6b. All the Raman modes exhibit preferential direction with twofold rotation symmetry characteristic for anisotropic crystals such as black phosphorus,[37] titanium trisulfide [38] or rhenium dichalcogenides.[39] The angle dependence of the intensity of the Raman modes can be used to determine the orientation of the crystal axes.

This is schematically presented in the inset of Figure 6a, where an optical micrograph of the flake together with arrows indicating the directions along which particular Raman modes reach maximum intensity. A dashed line indicates the direction $\theta = 0^\circ$ parallel to the ripples. The modes around 189 cm$^{-1}$, 263 cm$^{-1}$ and 322 cm$^{-1}$ reach maximum intensity when light is polarized perpendicular to the ripple direction, providing an additional tool to determine crystal lattice orientation.

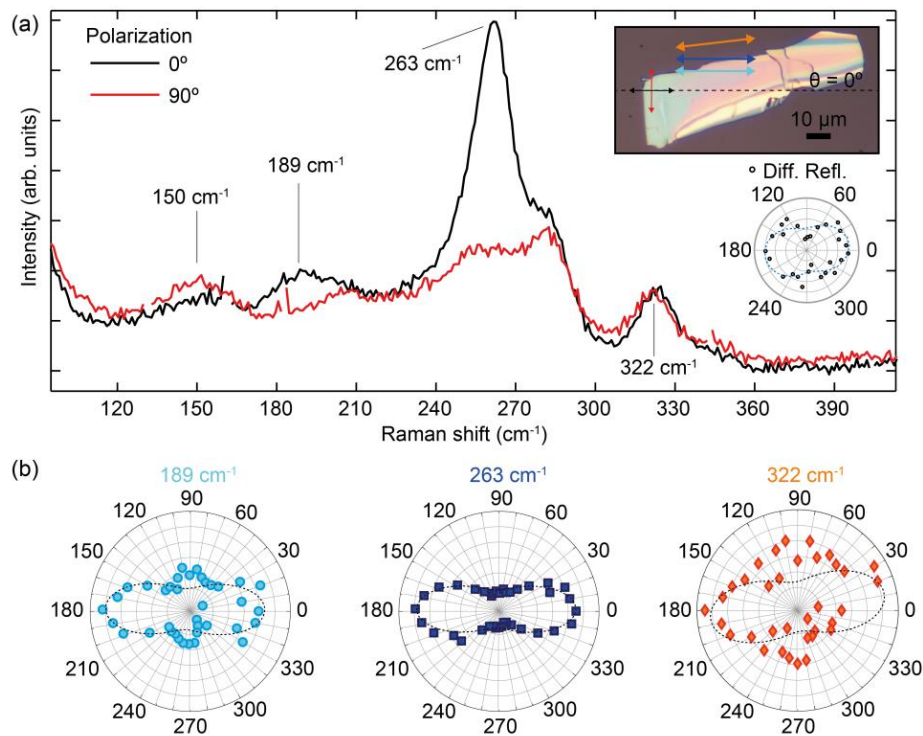

**Fig. 6. Anisotropic Raman spectra.** (A) Polarized Raman spectra with characteristic Raman modes from the Q and H phases. Inset: optical micrograph of the flake used for Raman measurements. The black dashed line marks the direction $\theta = 0^\circ$, parallel to the ripples. Arrows indicate the directions along which the intensity of the characteristic Raman modes is the highest. (B) Polar plots showing the area of the fitted peaks as a function of $\theta$ for three characteristic modes.

## Conclusions

In summary, we studied a naturally occurring van der Waals superlattice (franckeite) that clearly illustrates how new properties, absent in the individual constituent layers, emerge in the ensemble system. In particular, franckeite shows a periodic ripple structure in the TEM images that we could reproduce by our analytical continuum elasticity model. From the modelling we learnt that a balance between the van der Waals medi-ated adhesion and the young/shear modulus of the individual layers yields not only an out of plane deformation (ripple) but also to a simultaneous in-plane strain modulation. We further studied how this strong anisotropy, not present in the individual stacked constituents of the superlattice, affects the electrical and optical properties. We found that the conductance parallel to the ripples is twice as large as in the perpendicular direction. We believe that this anisotropic electrical conductance is the reason for the linear dichroism that we also observed on these samples: the optical absorption is larger for incoming light with linear polarization parallel to the ripples. We also observe intensity changes of characteristic Raman modes of the Q and H phase for different orientations of the excitation and scattered light polarization relative to the ripples. This further confirms the presence of optical in-plane anisotropy of franckeite which breaks the initial lattice symmetry of individual layers.

## Materials and Methods

**Materials.** Bulk natural franckeite mineral crystals (from San José mine, Oruro, Bolivia) have been used to extract thin flakes. The bulk crystals have been exfoliated with Nitto tape (224 SPV) onto a Gel-film® (WF x4 6.0 mil, by Gel-pak) substrate.

**TEM.** HRTEM observations were carried out in an aberration-corrected JEOL JEM-GRAND ARM300cF operated at 120kV and equipped with a cold field emission gun and a fast Gatan OneView camera. Strain analysis was performed by the geometric phase analysis (GPA) method [40] using the FRWRtools plugin for Digital Micrograph (www.physics.hu-berlin.de/en/sem/software/software_frwrtools).

**Optical microscopy.** Optical microscopy images have been acquired with an AM Scope BA MET310-T upright metallurgical microscope equipped with an AM Scope MU1803 camera with 18 megapixels. The trinocular of the microscope has been modified to connect it to a fiber-coupled Thorlabs spectrometer (part number: CCS200/M) to perform reflection and transmittance spectral measurements. [26]

**Fabrication of franckeite nano-devices.** Franckeite flakes are isolated from the bulk mineral by mechanical exfoliation with Nitto SPV224 tape and subsequently transferred to a $SiO_2$/Si substrate with an all dry transfer method.[22] We use standard e-beam lithography, e-gun metallization and lift-off procedures to define the metallic contacts (5 nm Ti/50 nm Au).

**Characterization of nano-devices.** The electrical characterization is performed in a *Lakeshore Cryogenics* probe station at room temperature in vacuum ($<10^{-5}$ mbar) using home-build source-meter unit electronics

**Raman spectroscopy.** Raman spectra were measured in two polarization configurations in a home-built setup. A 532 nm line of a CW solid state laser at 2 mW power was used for excitation through a 50x magnification long working distance microscope objective (NA=0.55). The linear polarization direction in sample plane was controlled by a rotating half-waveplate in the excitation beam and spectra were acquired in two detection polarization configurations: parallel (copolarization) and perpendicular (cross-polarization) to the polarization of excitation beam. The signal was dispersed by a 50 cm long spectrometer equipped with a 1800 grooves per millimeter diffraction grating and detected by a nitrogen cooled CCD camera. The measurements were carried out at room temperature in ambient conditions.

**Supporting Information**

Available at: https://pubs.acs.org/doi/abs/10.1021/acs.nanolett.9b04536

**Acknowledgements**

Electron microscopy observations were carried out at the Centro Nacional de Microscopia Electronica, CNME-UCM. This project has received funding from the European Research Council (ERC) under the European Union's Horizon 2020 research and innovation programme (grant agreement n° 755655, ERC-StG 2017 project 2D-TOPSENSE). ACG acknowledge funding from the EU Graphene Flagship funding (Grant Graphene Core 2, 785219). RF acknowledges support from the Spanish Ministry of Economy, Industry and Competitiveness through a Juan de la Cierva-formación fellowship 2017 FJCI-2017-32919. G.S-S acknowledges financial support from MINECO (Juan de la Cierva 2015 program, FJCI-2015-25427). PSJ acknowledges support from MINECO (FIS2015-65706-P).

**Competing Interests**

The authors declare no competing financial interests.

**Funding**

European Research Council (ERC): European Union's Horizon 2020 research and innovation programme (grant agreement n° 755655, ERC-StG 2017 project 2D-TOPSENSE)

EU Graphene Flagship funding (Grant Graphene Core 2, 785219)

Spanish Ministry of Economy, Industry and Competitiveness: Juan de la Cierva-formación 2017 FJCI-2017-32919

MINECO: Juan de la Cierva 2015 program FJCI-2015-25427 and FIS2015-65706-P

**Author Contribution**